# Repetition-Rate-Difference Tunable Dual-Comb Fiber Laser Using Bidirectional Lyot filtering

YUANJUN ZHU[1,2], BOWEN LIU[4]*, MAOLIN DAI[2,3], YIFAN MA[2,3], SHINJI YAMASHITA[2,3], TAKASUMI TANABE[4] AND SZE YUN SET[1,2]*

[1] *Department of Advanced Interdisciplinary Studies, Graduate School of Engineering, The University of Tokyo, Meguro-ku 153-8904, Japan.*
[2] *Research Center for Advanced Science and Technology, The University of Tokyo, 4-6-1 Komaba, Meguro-ku, Tokyo 153-8904, Japan.*
[3] *Department of Electrical Engineering and Information Systems, Graduate School of Engineering, The University of Tokyo, Bunkyo-ku 113-8656, Japan.*
[4] *Department of Electronics and Electrical Engineering, Faculty of Science and Technology, Keio University, 3-14-1, Hiyoshi, Kohoku-ku, Yokohama 223-8522, Japan.*
*\* b.liu@phot.elec.keio.ac.jp and set@cntp.t.u-tokyo.ac.jp*

**Abstract:** Single cavity dual-comb fiber lasers adopting different multiplexing configurations are benefited from the natures of common-mode noise suppression and superior coherence. Particularly, repetition-rate tunable dual-combs enable non-ambiguous ranging and aliasing-free spectroscopy. However, their sampling rate and spectral resolution is severely restricted by the mechanical delay lines. In a previous work, as rapid as 500 kHz/s tuning rate was realized to address this issue, while the minimum comb frequency difference remained large under the inaccuracy of mechanical DLL. In this work, a dual-comb prototype incorporated with a thermally controlled bidirectional lyot filter is demonstrated with 870-times enhanced tuning precision compared with mechanical schemes. Linear correlation between temperature and repetition-rate-difference of this tuning mechanism is revealed. We achieve a tuning efficiency of 4.4 Hz/°C and a control accuracy of 0.44 Hz/K, denoting a significant advance in operating Hz-scale differential comb lines. This design offers an optimal playground for extending non-ambiguous distance in dead-zone-free dual-comb ranging and eliminating aliasing in spectroscopy.

## 1. Introduction

Dual-comb laser systems have significantly advanced optical measurement techniques, enabling rapid, high-resolution spectroscopy, precision distance measurement, and advanced optical communication [1–4]. Conventional designs typically rely on two independent mode-locked lasers with slightly different repetition rates. However, maintaining mutual coherence between separate cavities requires complex stabilization and phase correction, substantially increasing system complexity. In contrast, single-cavity architectures inherently provide a shared optical path that suppresses common-mode noise and simplifies coherence management, offering a more practical and robust platform for dual-comb applications. Such systems eliminate the need for inter-cavity synchronization electronics. Various multiplexing approaches, including wavelength-division [5-9], bidirectional lasing [10-12], polarization multiplexing [13-14], and dual-branch configurations [15-17] have demonstrated the potential for robust, free-running dual-comb sources with reduced optical and electronic complexity.

Despite these advances, repetition-rate tunability is critical for dual-comb systems targeting alias-free ranging, extended non-ambiguity distance, and eliminating dead zones in real time [18-20]. In our previous work, we demonstrated a single-cavity, polarization-multiplexed dual-comb fiber laser incorporating a motorized delay line, achieving a tuning speed of approximately 500 kHz/s [21]. Optimization reduced the repetition rate difference to 184.5 Hz. However, the open-loop nature of the mechanical DLL (delay line) limits precise and repeatable reduction of the comb spacing, with an absolute displacement uncertainty of ±70 μm translating to a repetition rate error of ~383 Hz. This presents a fundamental barrier for precision applications unless closed-loop feedback is used to stabilize the cavity and suppress environmental drift.

To address this limitation, we incorporate a temperature-controlled Lyot filter [22-26] into our previous polarization-multiplexed design, enabling continuous and high-resolution repetition rate tuning via birefringence control. Unlike earlier birefringence-based methods relying on spliced PMF segments with limited tunability [14], our approach enables stable and repeatable tuning at the Hertz level. As a result, this configuration achieves a minimum repetition rate difference of 85 Hz with a tuning uncertainty of only 0.44 Hz, which is approximately 870 times more precise than our previous mechanical DLL only implementation. The demonstrated architecture ensures precise and repeatable tuning of the repetition rate difference in a single-cavity setup, addressing key requirements for aliasing-free dual-comb spectroscopy and extended-range distance metrology without dead zones.

## 2. Experimental principle

The repetition rate difference between two combs can be directly modulated by introducing a cavity length mismatch. The approximate relationship is given by:

$$\Delta f_{rep} \approx \frac{f_{rep}^2}{c} \times \Delta L \tag{1}$$

where $f_{rep}$ is the repetition rate, c is the speed of light in vacuum, and $\Delta L$ is the path length difference between two arms. In polarization-multiplexed dual-comb systems, repetition rate differences are often introduced by incorporating PMFs with controlled orientation and length mismatch. While this method allows frequency offsets in the kilohertz range, it suffers from poor tunability and limited repeatability due to strict fabrication tolerances and the static nature of the fiber layout.

To this end, we introduce a thermally controlled Lyot filter that links the filtered wavelength to the birefringence of a polarization-maintaining fiber segment. As the temperature increases, the birefringence B decreases, resulting in a blueshift of the filter's transmission peak. In the anomalous dispersion regime, where the C- and L-bands reside, the group index $n_g$ gradually decreases with decreasing wavelength [27]. As a result, the blueshift leads to a reduction in group delay, and consequently, an increase in the repetition rate. Therefore, thermal tuning indirectly modifies the round-trip delay by adjusting the lasing wavelength through birefringence. At the same time, as birefringence decreases, the difference between the group indices of the two polarization axes also diminishes, enabling precise and continuous control of the repetition rate difference via temperature.

The Lyot-filter-tuned laser is developed based on a previously proposed polarization-multiplexed configuration. Two design options are available for integrating the Lyot filter: a single-branch filtering scheme, where the filter is placed in only one polarization branch, and a main-cavity filtering scheme, where the filter is embedded within the shared main cavity. As illustrated in Fig. 1, these two configurations introduce similar yet distinct tuning mechanisms, resulting in different tuning characteristics.

In the single-branch filtering scheme shown in Fig. 1(a), adjusting the Lyot filter tunes only the lasing wavelength of the slow-axis branch. The difference in slow- and fast-axis group indices determines repetition rate difference. Consequently, the repetition rate difference is determined primarily by changes in the group index of that single branch which is the change in slow-axis group index, while the fast-axis remains unaffected.

In contrast, in the main-cavity filtering scheme shown in Fig. 1(b), both slow- and fast-axis combs experience identical spectral filtering. Thus, tuning the Lyot filter simultaneously modifies the lasing wavelengths and hence the group indices of both polarization axes. This approach provides three clear advantages: (i) a more precise tuning coefficient resulting from simultaneous contributions of both axes, (ii) enhanced robustness due to common-mode noise suppression, and (iii) improved coherence since both combs share the same filtered spectral center. For these reasons, the main-cavity filtering scheme is adopted in our system.

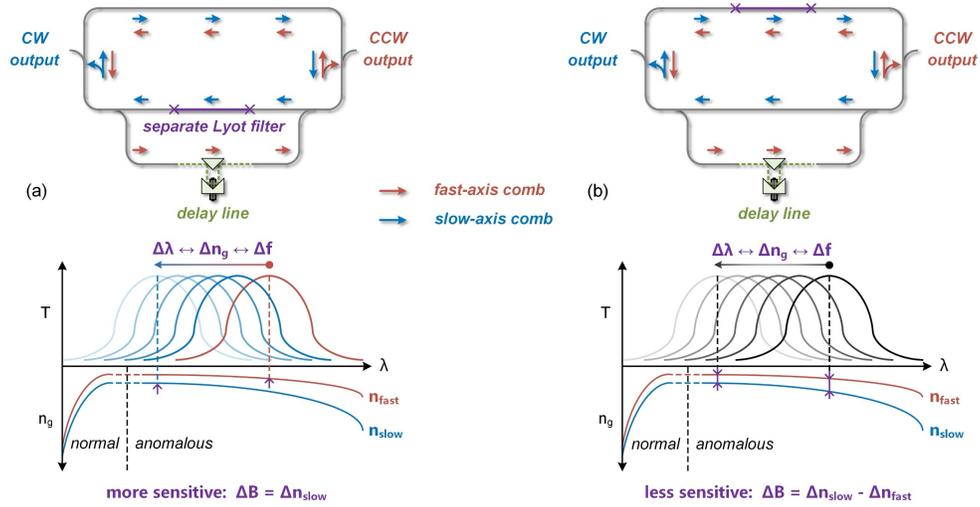

Fig. 1. schematic diagram of the proposed fiber laser: (a) Single-branch filtering scheme via inserting Lyot-filter in one of the branches; (b) Main-cavity filtering scheme via inserting a Lyot-filter in main cavity.

## 3. Experiment setup

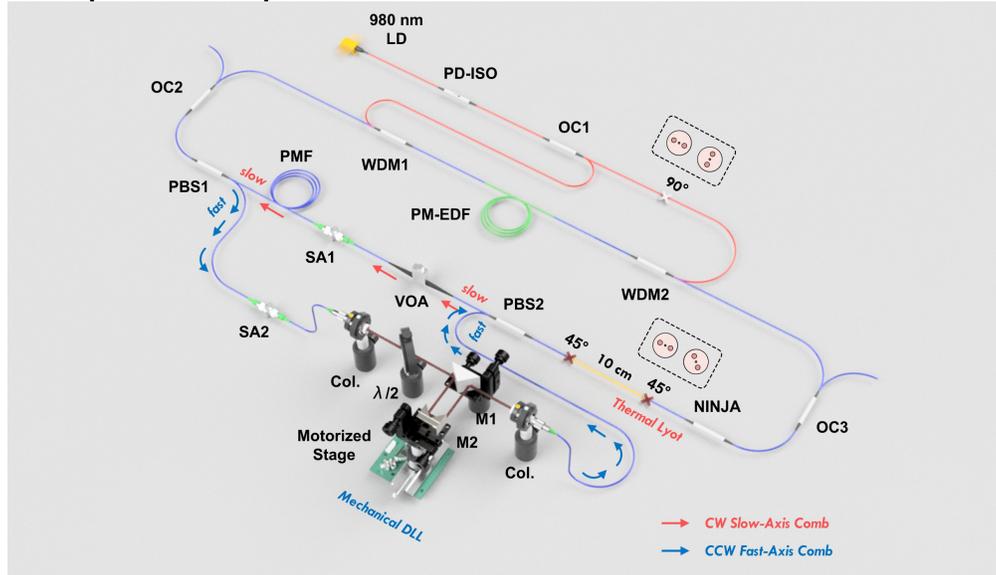

**Fig. 2.** Schematic diagram of the bidirectional polarization-multiplexed lyot filtering dual-comb fiber laser: WDM (wavelength division multiplexer); PM-EDF (polarization-maintaining erbium-doped fiber); OC (optical coupler); PBS (polarizing beam splitter); SA (saturable absorber); VOA (variable optical attenuator); NINJA (nonreciprocal isolating novel juxtaposition apparatus); LD (laser diode); PD-ISO (polarization dependent isolator); M (mirror), COL (collimator).

The schematic of the improved bidirectional polarization-multiplexed dual-comb laser is shown in Fig. 2. This cavity is based on the previous work in our group. [21] The gain medium consists of a 1.4 m long polarization-maintaining erbium-doped fiber (PM-EDF: PM-ESF-7/125) with a mode field diameter of 8.8 ± 1.0 µm, a birefringence of 3.5 × 10$^{-4}$, and a core absorption of 55 ± 5 dB at 1530 nm. A 980-nm laser diode pumps the PM-EDF in both directions at the same time by two wavelength division multiplexers (WDMs). Importantly, the pigtail between the pump source and WDM2 is spliced at 90° such that the pump light from WDM1 is propagated along the slow axis and the light from WDM2 along the fast axis.

Mode-locking is achieved by using two carbon nanotube (CNT)-based saturable absorbers (SAs), named SA1 and SA2, whose performance characteristics are comprehensively analyzed in reference [18]. Bidirectional polarization-orthogonal paths within the cavity are established using two polarizing beam splitters (PBS1 and PBS2) and a nonreciprocal isolating novel juxtaposition apparatus (NINJA) [28], a polarization-dependent isolator that allows light with a specific linear polarization to propagate in the designated direction. As a result, clockwise (CW) pulses (blue path) travel along the slow axis of the PMF and are extracted via optical coupler 1 (OC1). In contrast, counter clockwise (CCW) pulses (red path) propagate along the fast axis, pass through the motorized delay line (THORLABS ELL20/M), and exit via OC2. A 10-cm PMF spliced at 45° is placed between PBS2 and NINJA, enabling bidirectional polarization-multiplexed Lyot filtering. This configuration allows for simultaneous wavelength tuning of the two directional lasers via a temperature controller. The pigtails of all passive fiber components are made of standard PMF, and the total cavity length is approximately 7 m.

By placing the Lyot filter in the shared section of the cavity, both the CW and CCW pulses experience common spectral filtering. In our design, the CW pulses propagate along the fast axis, while the CCW pulses propagate along the slow axis of the PMF. The 45° splicing at both ends of the PMF ensures polarization rotation and symmetric filtering in both directions. Temperature fluctuations cause changes in birefringence in the PMF, which will in turn introduce an asymmetry in the group index over the two axes. This asymmetry generates fluctuations in the repetition rate of the two optical combs. The main-cavity filtering Lyot filtering technique ensures that the two combs share a common filtered spectral center, thus enabling mutual coherence and enabling the precise control of differences in repetition rates.

The laser output is characterized using the following measurement instruments: an optical spectrum analyzer (YOKOGAWA 6375) for spectral analysis, a 200 MHz oscilloscope (RIGOL DS2202A) in combination with a 1 GHz photoreceiver (NEW FOCUS 1611) for temporal waveform acquisition, a radio frequency (RF) spectrum analyzer (Rohde & Schwarz FPC1500) for RF spectrum analysis.

## 4. Experimental results

The wavelength tuning performance of the slow-axis comb (CW direction) is presented in Fig. 3(a), with the corresponding repetition rate variation shown in Fig. 3(b). As the Lyot filter temperature increases from 35 °C to 57 °C, the central wavelength exhibits a continuous blueshift from 1566.02 nm to 1542.24 nm, covering a tuning range of 24 nm. The optical spectra remain stable in shape and bandwidth throughout the range, indicating robust lasing behavior. Simultaneously, the RF repetition rate increases from approximately 29,828,060 Hz to 29,830,450 Hz, resulting in a total tuning span of ~2.4 kHz with a high signal-to-noise ratio (SNR) of up to 80 dB.

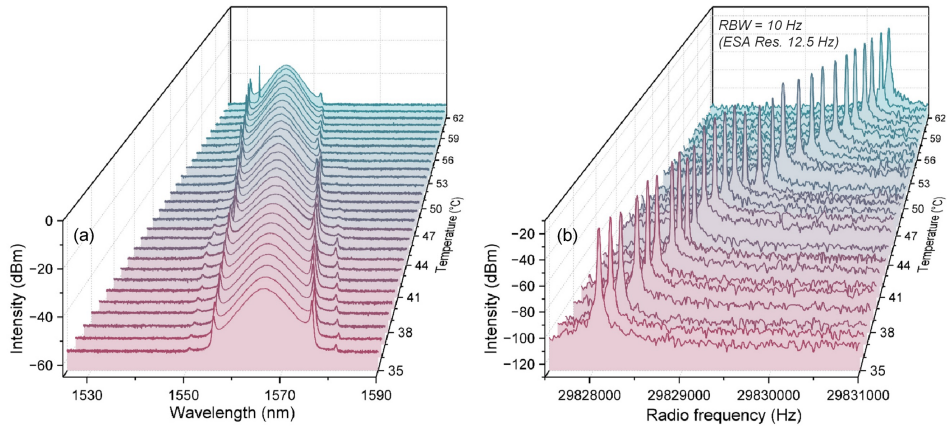

**Fig. 3.** Wavelength tuning and accompanying repetition rate tuning of slow-axis comb (a) Optical spectrum; (b) RF spectrum.

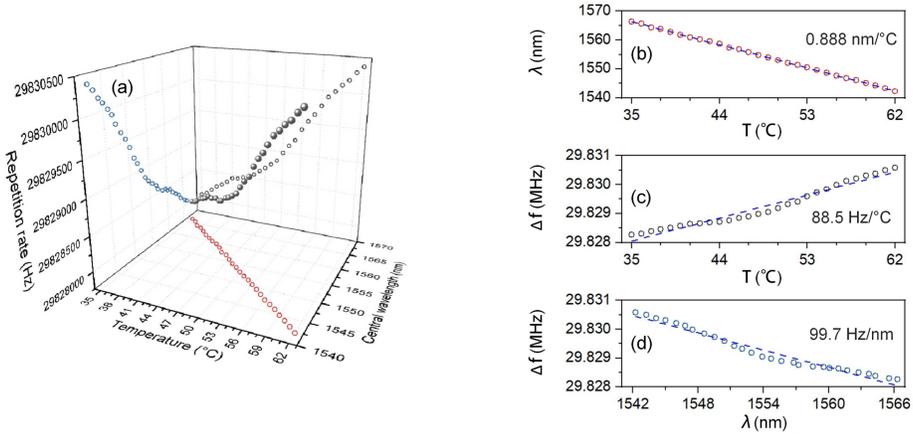

**Fig. 4.** (a) 3D plot showing the dependence of repetition rate, temperature, and central wavelength, (b) Linear wavelength shift with temperature, (c) Repetition rate variation with temperature, (d) Dependence of repetition rate on wavelength for slow-axis comb.

In addition to the slow-axis output, the fast-axis comb (CCW direction) also exhibits wavelength and repetition rate tuning under thermal modulation of the Lyot filter. The optical and RF spectra are shown in Figs. 5(a) and 5(b), respectively. Similar to the slow-axis case, the optical spectra remain stable in shape and bandwidth, with a smooth transition across the tuning range. Occasional continuous wave components are observed at certain wavelengths, but they do not affect the overall comb stability. The central wavelength shifts from 1566.27 nm to 1542.05 nm, yielding a tuning span of 24.2 nm, nearly identical to that of the slow-axis comb. Correspondingly, the RF repetition rate increases from approximately 29,828,260 Hz to 29,830,550 Hz, covering a range of about 2.3 kHz. Throughout this process, the RF peaks retain a signal-to-noise ratio exceeding 70 dB. The slight difference in tuning range between the two axes (about 100 Hz) reflects the intrinsic birefringence asymmetry in the PMF and its distinct group index response for each polarization mode. The above results affirm that the main-cavity Lyot filtering method enables continuous, high-resolution, and low-noise tuning of the repetition rates of the two frequency combs.

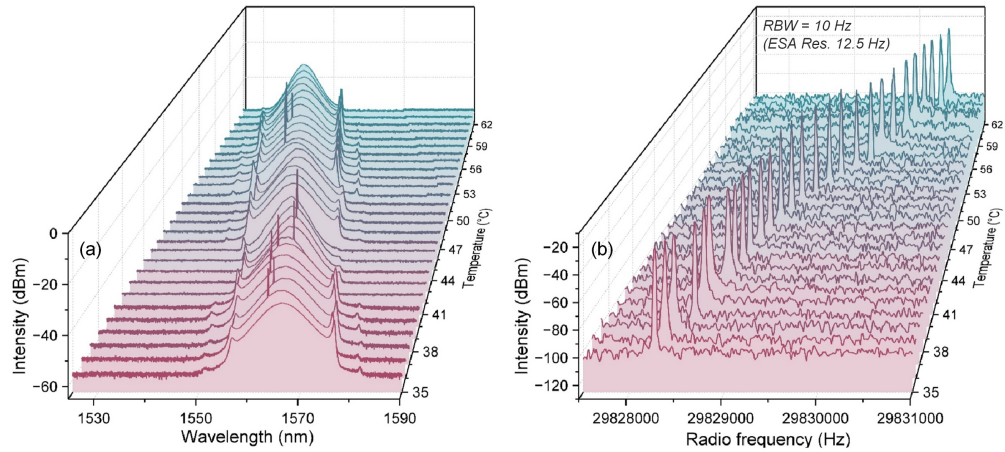

**Fig. 5.** Wavelength tuning and accompanying repetition rate tuning of fast-axis comb (a) Optical spectrum; (b) RF spectrum (span 10 kHz, RBW 10 Hz).

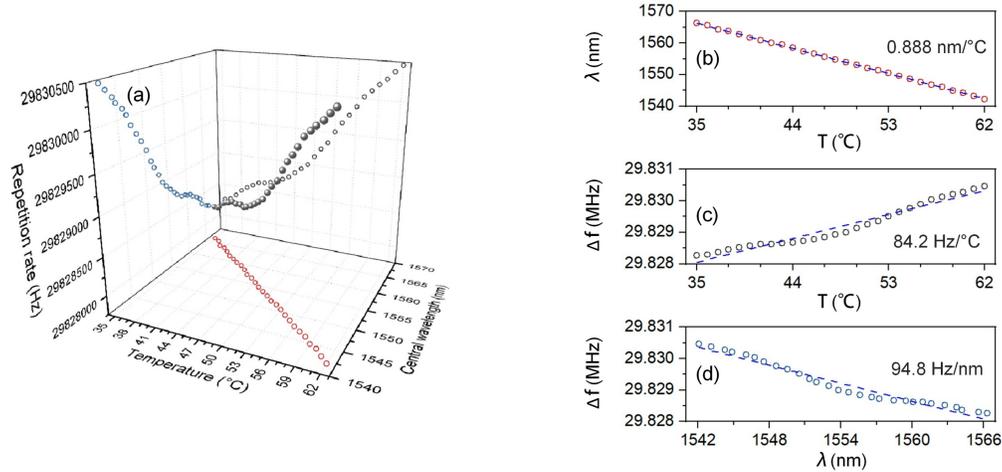

**Fig. 6.** (a) 3D plot showing the dependence of repetition rate, temperature, and central wavelength, (b) Linear wavelength shift with temperature, (c) Repetition rate variation with temperature, (d) Dependence of repetition rate on wavelength for fast-axis comb.

Tuning characteristics are summarized in Figs. 4 and 6. Linear fits yield thermal coefficients of 0.888 nm/°C for wavelength and 88.5 Hz/°C (slow axis) and 84.2 Hz/°C (fast axis) for repetition rate. These correspond to repetition rate sensitivities of ~99.7 Hz/nm and 94.8 Hz/nm, respectively. The slope mismatch enables sub-Hz control of the repetition rate difference, offering ~20-fold higher precision compared to single-axis tuning.

The repetition rate difference evolution is plotted in Fig. 7. As the temperature increases from 35 °C to 62 °C, the difference frequency reduces from ~210 Hz to 85 Hz, which is 100 Hz smaller than in PZT delay line-only systems. The differential tuning slope is ~4.4 Hz/°C or 5.0 Hz/nm. Considering a temperature control uncertainty of ±0.1 °C, the tuning error in repetition rate difference is estimated at ±0.44 Hz. Compared to the ±383 Hz error of mechanical delay lines [21], this reflects an ~871-fold improvement. The main-cavity filtering method also achieves ~20-fold higher precision than single-branch filtering, with improved stability and repeatability.

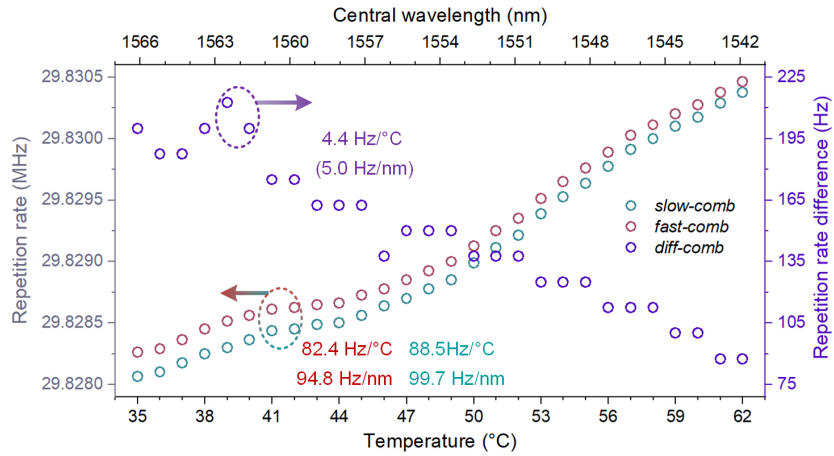

**Fig. 7.** Performance of repetition rate tuning by the main-cavity filtering dual-comb laser.

## 5. Discussions

To better understand the mechanism underlying temperature-dependent repetition rate tuning, we conducted numerical simulations based on the models in Refs. [29-31]. The simulations evaluated how birefringence and modal indices in PMFs evolve with temperature and influence the laser's repetition rate. In our setup, the central wavelength shifts from 1566 nm to 1543 nm as the temperature rises from 35 °C to 62 °C (a 27-K change). Figure 8 shows the simulated Lyot filter transmission versus temperature and wavelength, revealing a spectral shift due to thermally induced birefringence variation. Overlaid curves indicate the transmission peak shifts from 1566 nm to 1542 nm, while the birefringence decreases from $3.524 \times 10^{-4}$ to $3.470 \times 10^{-4}$. This corresponds to a tuning rate of $-1.998 \times 10^{-7}$ K$^{-1}$, consistent with theoretical predictions [29]. The negative sign confirms the expected decrease in birefringence with rising temperature.

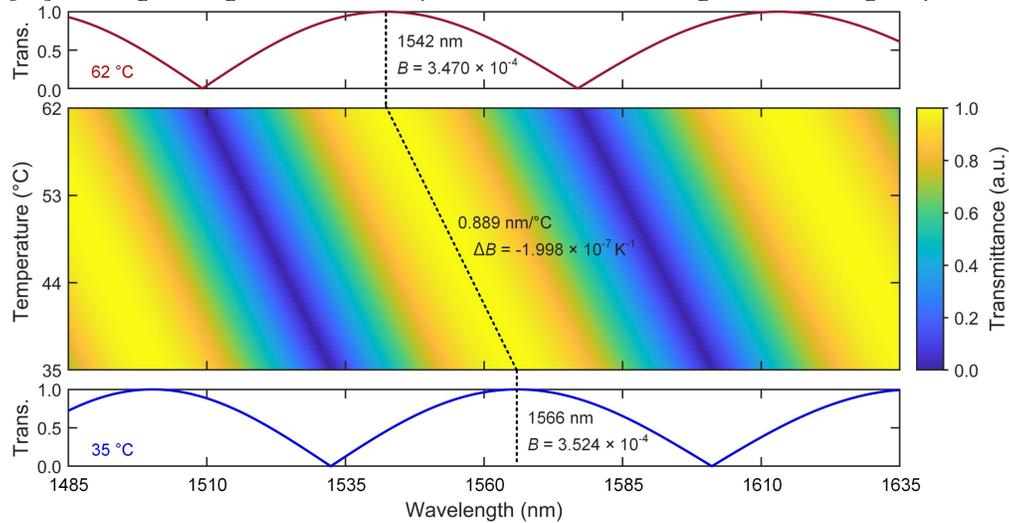

**Fig. 8.** Simulation results of wavelength tuning and birefringence change in the repetition-rate-tunable laser.

According to the first-order differentiation of the repetition rate, the relationship can be quantitatively described as:

$$\Delta f_{rep} = f_{rep} \times \frac{\Delta n_{slow}}{n_{slow}} \qquad (2)$$

Assuming that the slow-axis group index is 1.467 at 1550 nm, its rate of change is thereby calculated as $-41.894\times10^{-7}$ K$^{-1}$. Then, the one of fiber fast axis is converted to be $-39.896\times10^{-7}$ K$^{-1}$. The negative sign indicates the decreasing trend as the temperature increases.

Figure 9(a) illustrates the simulated evolution of these indices. As temperature rises, the central lasing wavelength undergoes a blueshift. Concurrently, the birefringence decreases due to thermal relaxation of fiber stress, while the individual modal indices increase. This observation is consistent with the known slow increase of group index in the anomalous dispersion regime. The corresponding simulated changes in repetition rate and their differential values are shown in Fig. 9(b), which exhibit trends opposite to those of the modal indices. Compared to experimental measurements, the simulated tuning coefficients are slightly lower: 85.18 Hz/°C and 81.16 Hz/°C for the slow- and fast-axis combs, respectively, with a difference of 4.02 Hz/°C. This small discrepancy may arise from deviations between the assumed group index (1.467) and its actual value.

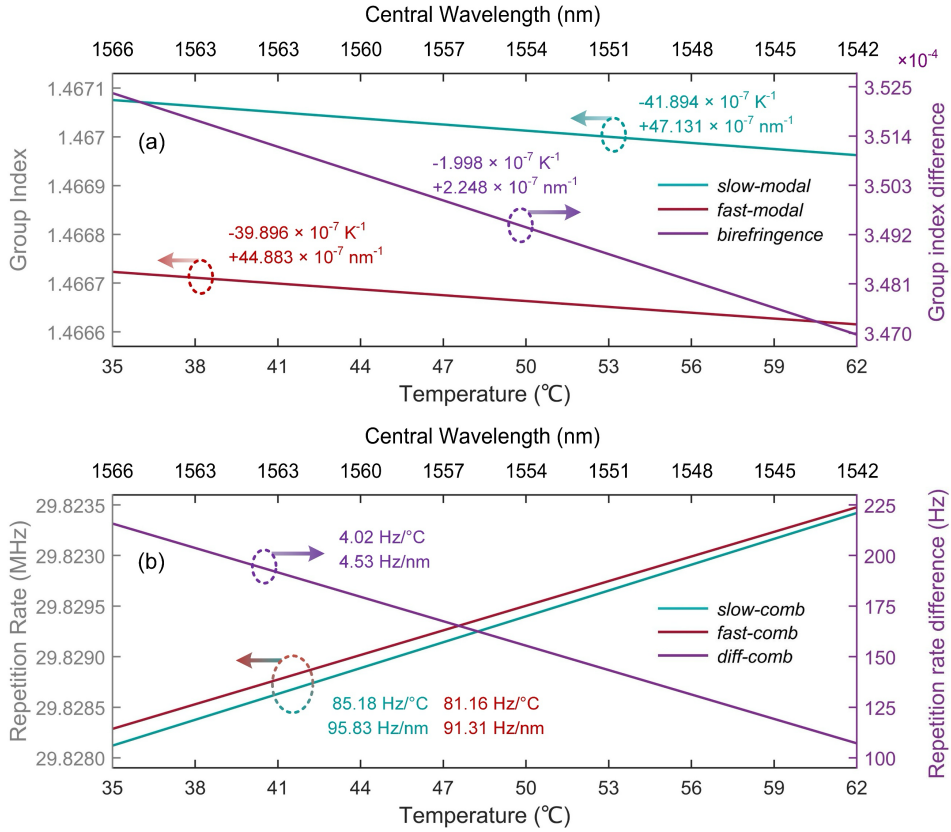

**Fig. 9.** Simulation results of (a) polarization-dependent modal index change and (b) related repetition rate alteration in the repetition-rate-tunable laser.

## 6. CONCLUSION

In summary, we present a precision-enhanced dual-comb fiber laser, where a temperature-controlled lyot filter is incorporated as the key component into a previously developed mechanical DLL driven bidirectional PM configuration. Common lyot filtering scheme, which depends on the relative change between slow and fast modal group indices, is applied to offer a greatly improved tuning precision. The bidirectional lyot filter converts the heating temperature linearly into the shifts of central wavelengths on both slow and fast-axis combs, which are linked with group indices' relative variations, and thus prompts the tuning of repetition-rate-difference. This mechanism contributes to the 4.4 Hz/K tuning efficiency in our experiment, indicating a significant advance for operating dual-comb measurement with an extremely high frequency resolution that requires Hz-scale differential comb teeth density. Given that the minimum temperature division of the heater is 0.1 K, we obtain a rough estimation of tuning uncertainty of 0.44 Hz/K. This is approximately 20 times better than separate lyot filtering method, and furthermore, achieves ~870-time improvement in contrast to mechanical DLL previously reported [21]. The results demonstrate an optimal playground for extending the non-ambiguity range in dual-comb ranging and resolving aliasing limitations in broadband spectroscopy, with both excellent tunability and accuracy.

**Acknowledgment.** Dr. S. Y. Set thanks Mr. Hideru Sato and Mr. Raymond Chen for their kind personal donation for this research work.